\documentclass[shortnote,twocolumn]{jpsj2}
\usepackage{graphicx}

\def\runtitle{a sample}
\def\runauthor{Masayuki \textsc{Yamamoto}, Tomi \textsc{Ohtsuki } 
and Keith \textsc{Slevin}}

\title{%
Effects of magnetic field applied on leads
}

\author{%
Masayuki \textsc{Yamamoto}\thanks{E-mail:masayu-y@sophia.ac.jp},
Tomi \textsc{Ohtsuki} and Keith \textsc{Slevin}$^{1}$
}

\inst{%
Department of Physics, Sophia University, 7-1 Kioi-cho,
Chiyoda-ku, Tokyo 102-8554, Japan \\
$^1$Department of Physics, Graduate School of Science,
Osaka University, 1-1 Machikaneyama, Toyonaka, Osaka 560-0043, Japan \\
}

\recdate{\today}

\abst
{
}

\kword{%
Universality class, Transmission eigenvalue, Level statistics, Wigner ensemble
}

\begin{document}
\sloppy
\maketitle

\section{Introduction}
During the past decade, there has been considerable interest
in quantum transport phenomena in devices where the spin degree of freedom of
the electron plays a crucial role {\em e.g.} the spin-transistor\cite{Datta}.
Of particular interest are mesoscopic systems where the conductance
is sensitive not only to the sample's specific configuration but also to
the spin population in the leads attached to the sample.

One of the special characteristics of mesoscopic systems is
the fluctuating nature of transport coefficients
such as the conductance.
Such fluctuations do not depend on the details
of the sample and the strength of disorder but are usually 
determined by the symmetry and the dimensionality of the system.
The standard classification\cite{Beenakker} is detailed in Table.\ref{uni-table}  

\begin{table}
\caption{Universality class and corresponding symmetry of the system
and the properties of scattering matrix.}
\label{uni-table}
\begin{tabular}{lllll}
\hline
Universality class & \multicolumn{2}{l}{Symmetry} & 
Properties of S matrix \\
 & TRS & SRS & \\ 
\hline
Orthogonal & yes & yes & unitary symmetric \\
Symplectic & yes & no  & unitary self-dual \\
Unitary & no & irrelevant & unitary \\
\hline
\end{tabular}
\end{table}

The two relevant symmetries are
time reversal symmetry (TRS) and spin rotation symmetry (SRS).
TRS is broken by an applied magnetic field or by
magnetic scattering due to magnetic impurities or magnetic domain walls.
If TRS is broken, systems are classified as unitary,
regardless of whether or not SRS is broken.
The spin-orbit interaction breaks SRS but
preserves TRS, and in this case systems are classified as 
symplectic.

The relation between the Hamiltonian of the sample and its 
corresponding universality class is well understood.
However, the properties of mesoscopic systems that are measured
in transport experiments depend not only on the sample but also
on the leads through which currents flow in and out of the
sample and through which voltages are measured.

In this contribution we address the question of whether or not the universality 
class can change as a result of an asymmetric spin population in
the leads
i.e. when the number of up and down spin channels in the leads
becomes asymmetric as a result of Zeeman splitting.
We show that even if the sample is unchanged, the
universality class can change under certain conditions.

\section{Model}
We consider a two dimensional (2D) sample connected
to two electron reservoirs by ideal leads.
The 2D system is constructed in the $xy$-plane with the
$x$ direction taken as the direction of current flow. 
The sample region contains both a random potential and a spin-orbit
interaction described by the Ando Hamiltonian\cite{ando}
\begin{equation}
  {\cal H} = \sum_{i,\sigma}W_{i} c_{i\sigma}^{\dagger}c_{i\sigma}
 - \sum_{<i,j>,\sigma,\sigma'} V_{i \sigma,j\sigma'}
c_{i\sigma}^{\dagger}c_{j\sigma'}
\end{equation}
where
\begin{equation}
V_{i,i+\hat{x}} =
\left( 
\begin{array}{cc}
\cos \theta & \sin \theta \\
-\sin \theta & \cos \theta \\ 
\end{array}
\right)
\end{equation}
and
\begin{equation}
V_{i,i+\hat{y}} =
\left( 
\begin{array}{cc}
\cos \theta & i \sin \theta \\
i \sin \theta & \cos \theta \\ 
\end{array}
\right)
\end{equation}
Here $c_{i\sigma}^{\dagger} (c_{i\sigma})$ denotes
the creation (annihilation) 
operator of an electron on site $i$ with spin $\sigma$.
$W_{i}$ denotes the random potential on the site $i$,
distributed uniformly on $[-W/2,W/2]$.
The hopping is restricted to nearest neighbours.
The unit vector of the $x(y)$-direction is denoted by 
$\hat{x}(\hat{y})$.
The parameter $\theta$ denotes the strength of the spin-orbit interaction.
The strength of the hopping is taken to be the energy unit.

We impose the fixed boundary condition in the transverse direction $y$.
The transverse energy of the channel $(i,\sigma)$ in the lead 
$\varepsilon_{i}^\sigma$ is given by
\begin{equation}
\varepsilon_{i}^\sigma = -2\cos(\frac{i\pi}{n+1})-Z\sigma
\hspace{1cm} (i=1,2,\cdots,n)
\end{equation}
where $n$ denotes the number of sites in $y$-direction, and
$Z$ denotes the strength of Zeeman splitting and
$\sigma (=\pm 1)$ is the spin index. 
The channel $(i,\sigma)$ is a propagating mode if 
\mbox{$|E_{F} - \varepsilon_{i}^\sigma| < 2.0$},
$E_{F}$ being the Fermi energy.
For example, when we set $E_{F}=-1.1$ and $Z=1.0$,
the number of up (down) spin channels becomes 27 (15) for the sample of
width 30 sites.

In the following simulation, 
we set to be $W=2.0$, $\theta = 0, \pi/4$ and $E_{F} = -1.1$.
The system size is set to be $30 \times 30$ in units of the lattice
spacing.

\section{Results}
We extend the transfer matrix method \cite{pendry} to the case 
where there is a spin-orbit interaction in the sample
and the population of up and down spins in the leads is asymmetric.
We calculate the transmission matrix $t$ in the two terminal
configuration.
The population of up and down spins in one lead is always set to be
symmetric ($Z=0$), and that in the other lead is varied ($Z=0,1,2$).

To detect any change of universality class, we investigate 
the spacing distribution $P(s)$ where $s$ is
the interval between neighboring transmission
eigenvalue $\tau$ (the eigenvalues of $t t^\dagger$).
An ensemble of about $10^6$ samples is simulated
for the estimation of the level statistics.

Figure 1 shows the spacing distribution $P(s)$
of the transmission eigenvalue $\tau$
for the sample with a spin-orbit interaction.
In the absence of Zeeman splitting in the leads ($\circ$) 
$P(s)$ is close to the Wigner surmise for the
Gaussian symplectic ensemble (GSE).
On the other hand, with Zeeman splitting in the leads
($\bullet$) $P(s)$ is close to the surmise for
the Gaussian unitary ensemble(GUE).
The reason for this crossover is that the
asymmetry of the population of up and down spins in the lead destroys 
the self-dual property of the scattering matrix.

Figure 2 shows the spacing distribution of the transmission eigenvalue $\tau$
for the sample with $\theta=0$ i.e. no spin-orbit interaction.
In contrast to Fig. 1 the level statistics do not 
depend on the states of the leads and are well fitted by the
formula for the Gaussian orthogonal ensemble (GOE) .
We summarize these results in Table. \ref{lead-table}.

In summary, we have shown that the level statistics of the transmission
coefficient is sensitive to the asymmetry of the spin population in the
lead. We can thus control the conductance fluctuations without changing
the sample parameters. 
For example, the variance of conductance is suppressed by $1/2$
when we introduce the asymmetry of the spin population in a lead.
(Note that the GUE in the presence of the asymmetry
is spin nondegenerate, and the variance of conductance is
1/4  the spin degenerate value.)
This behavior is essentially different from that of the asymmetry of the
number of channels between the leads where the universality class is not
changed. 

\pagebreak

\begin{table}
\caption{Effects of Zeeman splitting in a lead}
\label{lead-table}
\begin{tabular}{lll}
\hline
SO in a sample & Zeeman in a lead & Universality class \\
\hline
yes & no & Symplectic \\
yes & yes & Unitary \\
no & no & Orthogonal \\
no & yes & Orthogonal \\
\hline
\end{tabular}
\end{table}

\begin{figure}
\begin{center}
\includegraphics[scale=0.35]{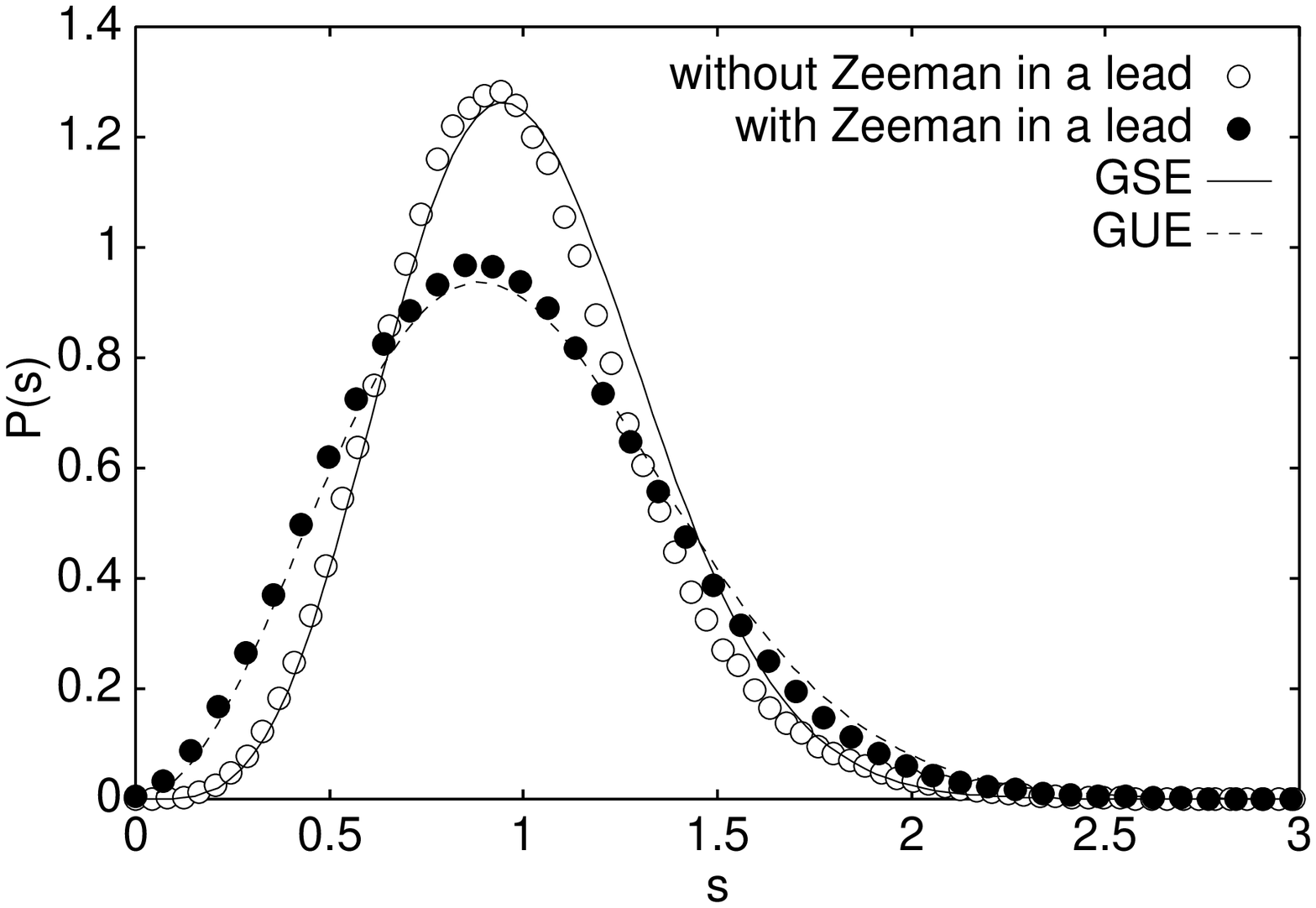}
\caption{Spacing distribution of $\tau$ for the sample with the spin-orbit
interaction.
We set $Z=2$.
The distribution without Zeeman splitting in a lead ($\circ$) fits to
the form of GSE 
while that with Zeeman ($\bullet$) fits to GUE.}
\end{center}
\end{figure}

\begin{figure}
\begin{center}
\includegraphics[scale=0.35]{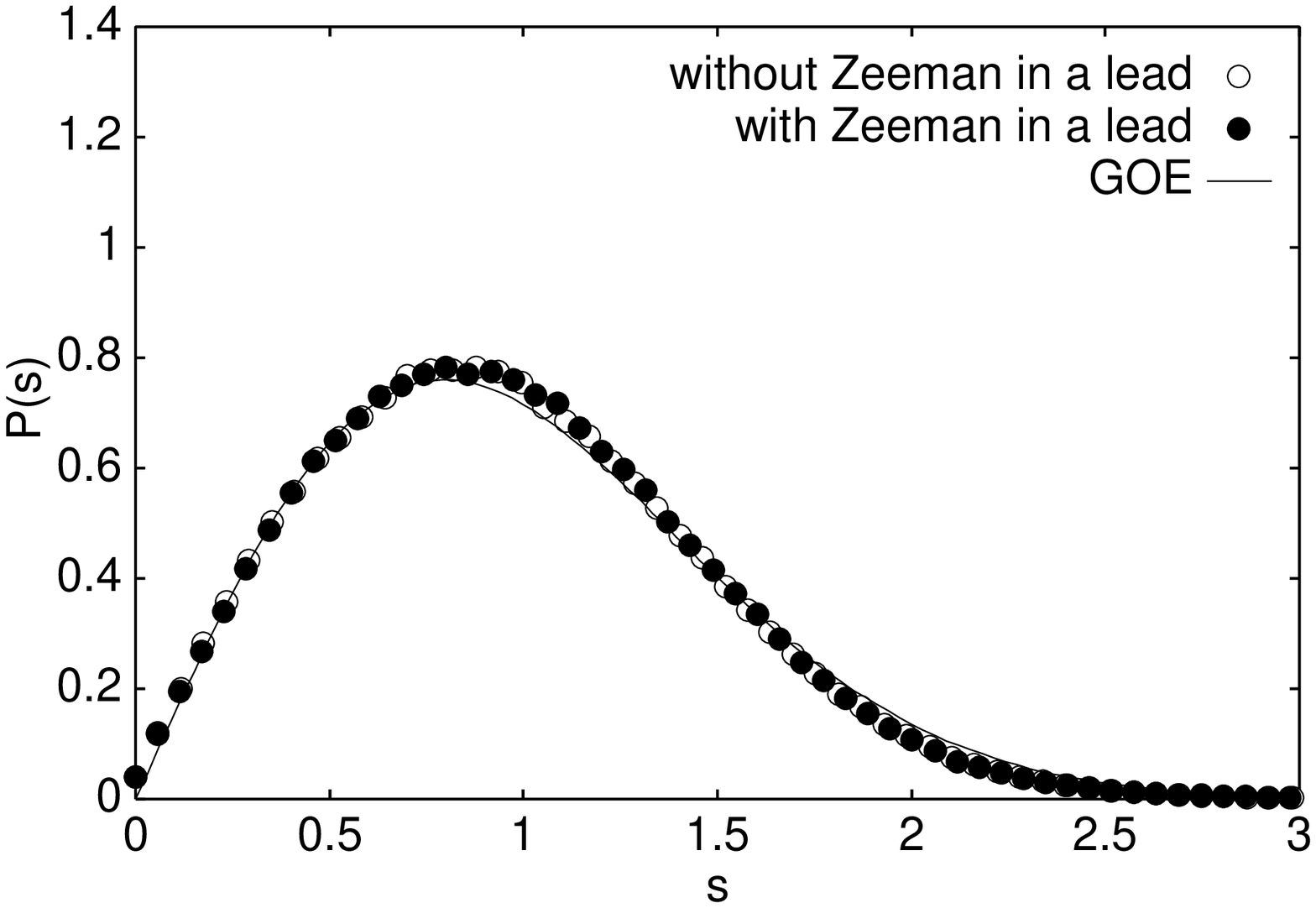}
\caption{Spacing distribution of $\tau$ for the sample without the spin-orbit
 interaction.
We set $Z=1$.
The distribution fits GOE regardless of the existence of Zeeman splitting in a lead.}
\end{center}
\end{figure}

\end{document}